# Structural and magnetic properties of $\varepsilon$-Fe$_{1-x}$Co$_x$Si thin films deposited via pulsed laser deposition

Ncholu Manyala[1], Balla D. Ngom[2,3], A. C. Beye[2], Remy Bucher[3]. Malik Maaza[3], Andre Strydom[4], Andrew Forbes[5], A. T. Charlie Johnson, Jr.[6], and J. F. DiTusa[7]

[1]*Department of Physics, Institute of Applied Materials, University of Pretoria, Pretoria, RSA*

[2]*Groupe de Laboratoires de Physique des Solid et Sciences des Matériaux, Faculté des Sciences et Techniques, Université Cheikh Anta Diop de Dakar, B. P. 25114 Dakar-Fann Dakar, Sénégal*

[3]*Nano-Sciences Laboratories, Materials Research Group, iThemba LABS, National Research Foundation, P.O. Box 722, Somerset West 7129, RSA*

[4]*Department of Physics, University of Johannesburg, Johannesburg, RSA*

[5]*CSIR National Laser Centre, Pretoria, RSA*

[6]*Department of Physics and Astronomy, University of Pennsylvania, Philadelphia, PA 19104, USA*

[7]*Department of Physics and Astronomy, Louisiana State University, Baton Rouge, Louisiana 70803, USA*

(May 15$^{th}$, 2009)

We report pulsed laser deposition synthesis and characterization of polycrystalline $Fe_{1-x}Co_xSi$ thin films on Si (111). X-ray diffraction, transmission electron, and atomic force microscopies reveal films to be dense, very smooth, and single phase with a cubic B20 crystal structure. Ferromagnetism with significant magnetic hysteresis is found for all films including nominally pure FeSi films in contrast to the very weak paramagnetism of bulk FeSi. For $Fe_{1-x}Co_xSi$ this signifies a change from helimagnetism in bulk, to ferromagnetism in thin films. These ferromagnetic thin films are promising as a magnetic-silicide/silicon system for polarized current production, manipulation, and detection.

$Fe_{1-x}Co_xSi$ has been recently identified as a silicon-based magnetic semiconductor as it spans the insulating, metallic, and spin polarized metallic regimes as a function of Co concentration[1,2]. The parent compound, $\varepsilon$-FeSi, is small band-gap insulator with a cubic B20 crystal structure[3-7]. $Fe_{1-x}Co_xSi$ is of interest because of its potential as a material for silicon based spintronics applications[1] where magnetic semiconducting films serve as injectors and collectors of spin-polarized currents in Si transistors. Films that have properties similar to that of bulk crystals and that form clean interfaces with Si are required for this technology. The ability to grow epitaxially registered silicides of Co and Ni on silicon substrates with atomically smooth interfaces[8,9] further motivates us to explore $Fe_{1-x}Co_xSi$ film growth. Synthesis of low dimensional FeSi is challenging due, in part, to the complex phase diagram of Fe and Si which includes five known bulk phases[10]. However, growth of free standing FeSi nanowires by chemical vapor deposition methods has been recently reported[11,12]. Most promising is a novel approach using single-source precursors as it allows $Fe_{1-x}Co_xSi$ and CoSi nanowire growth[13,14]. Syntheses of FeSi films employing diverse deposition methods, and either an Fe or a $\beta$-$FeSi_2$ target[15-20], result in mixed iron silicide phase films with $\beta$-$FeSi_2$ being a common impurity phase formed during post-deposition annealing[20]. Previous pulsed laser deposition (PLD) syntheses of FeSi on Si substrates held at temperatures ($T$) between $25 \leq T \leq 900$ °C report amorphous film growth[21,22].

Here, we report the synthesis of polycrystalline $Fe_{1-x}Co_xSi$, $0 \leq x \leq 0.3$, thin films, without measurable secondary phases, by PLD techniques. We demonstrate that these films have hysteretic magnetic ground states in contrast to the nearly history independent helimagnetism of bulk $Fe_{1-x}Co_xSi$. Our success in synthesizing crystalline films can be attributed to the careful attention paid to the PLD growth parameters.

Films were deposited on Si (111) substrates held at 450 $^0$C in vacuum ($10^{-6}$ Torr). Substrates were first cleaned ultrasonically with several organic solvents as well as with de-ionized water and etched in a 20% HF-solution. Substrates were left in air for a period of between 2 and 30 days prior to deposition. An excimer laser, wavelength of 248 nm, fluence of 4 J / cm$^2$, repetition rate of 10 Hz, and 30 ns pulse duration, was incident on arc melted[2] $Fe_{1-x}Co_xSi$ targets for between 5 and 30 minutes. (Cu-$k\alpha$) X-ray diffraction (XRD) patterns using an AXS Bruker diffractometer equipped with a position sensitive detector determined the crystalline structure. The surface morphology was characterized via scanning electron microscopy (SEM) and atomic force microscopy (AFM) while transmission electron microscopy (TEM) probed the cross-sectional microstructure of the films. Film stoichiometry, determined by Rutherford backscattering spectroscopy (RBS) and energy dispersive spectrometetry, TEM (EDS) and X-ray fluorescence SEM (EDX), agreed with the concentrations of our starting materials showing a transition metal to Si ratio of 1:1($\pm$0.5%) and a Fe to Co ratio within 2%. EDS line scans revealed Fe diffusion to 100 nm below the substrate surface consistent with previous reports[21,22]. The zero magnetic field, $H$, cooled (ZFC) and field cooled ($H>0$: FC) magnetization, $M$, of our films was measured with a SQUID magnetometer with the small background due to the substrate carefully subtracted. Preliminary 4-probe resistivity measurements were performed on 2 films, both of which displayed metallic behavior with a residual resistivity ratio of 1.7. A more complete treatment of the electrical transport is presently being carried out.

We have identified all discernable peaks in our XRD scans, see e.g. Fig. 1, as belonging to either the B20 structure of the films or the diamond structure of the Si substrate. No evidence for secondary impurity phases has been found in any of the scans indicating that the films are likely single phase. The lattice constants, $a$, of the films, presented along with bulk values as a function of $x$ in the inset[1,2], are larger than in bulk samples by ~0.3% at small $x$. The lattice 17% mismatch between the Si (111) substrate and the FeSi films may cause significant strain near the film-substrate interface. However, as our films are between 35 and 500 nm thick, changes in $a$ due to interfacial strain are not expected since it is typically relieved via crystalline defects within a few nm of the interface. Stoichiometric fluctuations have been ruled out as the cause of the expanded lattice by our RBS and EDX measurements.

High magnification TEM images of our films reveal sharp and continuous film-substrate interfaces as in Fig. 2a where the Si lattice in the lower right half and the FeSi lattice at the top are clearly defined. Although nanobeam diffraction (NBD) patterns displayed distinct diffraction spots, no single crystals were isolated despite the use of ~20 nm analytical area. The NBD, in agreement with XRD, shows multiple sets of spots derived from several crystals indicating poly-crystallinity. The AFM, Fig. 2b, and SEM images display almost featureless, extremely smooth (< 0.8 nm root-mean-square surface roughness) film surfaces that are free of clusters and islands with the exception of a small density of droplets common to PLD growth.

The magnetization and magnetic susceptibility, $\chi$, of a set of our films along with our corresponding high purity bulk samples[1,2] are shown in Figs. 3 and 4. The 50 Oe FC and ZFC $\chi$ versus $T$ for films with $x$ = 0, 0.2, and 0.3 as well as our FeSi single crystal at 1 kOe[23,24] is displayed in Fig. 3. As has been well documented, $\chi$ of crystalline and polycrystalline FeSi approaches zero at $T<100$ K in a manner consistent with an energy gap for electronic excitations of 60 meV[3,4,7]. In contrast, our FeSi film displays a $\chi$ similar to disordered ferromagnets with a significant history dependence. Our $Fe_{1-x}Co_xSi$ films display a similar $T$-dependence with a systematic increase in $\chi$ with $x$. The Curie temperature, $T_C$, of our films which we associate with the peak in ZFC $\chi$, along with those of bulk samples[1,2], is displayed in the inset of Fig. 3, where a much more subtle $x$ dependence is apparent. The magnetic state of our films at higher fields is shown in Fig. 4a where $M$ at $H=10$ kOe is displayed for the same films shown in Fig. 3, along with 2 polycrystalline pellets having the same $x$ [Ref. 2]. $M$ of the films with $x = 0.2$ and 0.3 is comparable to, but slightly larger than, the bulk $M$ at low-$T$ from previous experiments that indicated, and band structure calculations supported[25], a high carrier spin polarization[2]. $M$ of our films closely resembles that of prototypical weak itinerant magnets such as $Sc_3In$[26]. The gradual decay of $M$ at $T>T_C$ may indicate either a small Fermi energy allowing ferromagnetic fluctuations survive up to high $T$, or that there is local ordering in the films at $T>T_C$. The possible occurrence of clusters of ordered, strongly fluctuating, magnetic moments at $T>T_C$ is comparable to that in disordered ferromagnetic metals such as $Fe_{1-x}Co_xS$[27], and is further supported by the difference between ZFC and FC data of Fig. 3.

In addition to the differences noted above between bulk and film samples, we observe significant differences in the 5 K $M(H)$ displayed in Fig. 4b. The behavior of bulk $Fe_{1-x}Co_xSi$, representatively displayed for an $x = 0.15$ sample[2], is typical of a helimagnet in that $M(H)$ goes through zero at $H = 0$ and shows a small variation between the initial and subsequent $H$-sweeps[2]. In contrast, all our thin films, including our FeSi film, display behavior similar to standard ferromagnets with a coercive field of ~100 Oe and a saturated $M$ at higher $H$. The cause for the change from helimagnetism to ferromagnetism is not known, although pinning of $M$ by the surfaces of the film may play an important role.

Although our films with $x > 0$ resemble the bulk polycrystalline materials in that their low-$T$ saturated $M$'s approach full carrier polarization, our nominally pure film is vastly different from bulk FeSi. As we have noted, $a$ is 0.3% larger than for bulk FeSi and we point out that LDA band structure calculations for FeSi and isostructural FeGe, a helimagnetic metal ($a$ = 0.47 nm) and a $T_C$ of 280 K[28], predict an unusual sensitivity to changes in $a$ due to the small energy gap at the Fermi level[29]. For FeSi, band structure calculations using our experimentally determined $a$ predict a metallic and magnetic ground state[30] consistent with what we observe. A second cause may be the presence of small variations of the Fe to Si ratio. Although our RBS and EDX measurements confine the stoichiometry, they are averaged over large film areas so that local variations are possible. We do not consider the presence of Fe diffused into the substrate to contribute to $M$ since Fe ions that are well coordinated with Si have very small, ~zero, magnetic moments[31]. Future experiments are planned to explore the cause of the dramatic differences we measure in bulk and thin film samples as well as the possibility of relaxing the lattice constant toward the bulk value.

We have demonstrated that high-quality polycrystalline cubic B20 $Fe_{1-x}Co_xSi$ thin films can be successfully grown on Si (111) substrates with magnetic properties that are broadly similar to that of the high purity single and poly-crystalline bulk samples.

However, there are significant differences that these films display from bulk behavior, including a $T_C$ that is larger at small $x$ and a $M(H)$ at $T < T_C$ that is much better described as ferromagnetic than helimagnetic. Our nominally pure FeSi thin films display substantial differences with the behavior of bulk FeSi which we attribute to differences in $a$ and/or to local differences in stoichiometry. This report of the growth of crystalline $Fe_{1-x}Co_xSi$ in thin film form via laser ablation techniques is an important step towards the demonstration of injection of polarized currents in silicon via magnetic $Fe_{1-x}Co_xSi$ ($0.1 \leq x \leq 0.3$) / Si interfaces.

N. Manyala acknowledges support from the African Laser Centre, NanoAFNET, iThemba LABS, and the NRF of South Africa. The LRSM, U. of Pennsylvania, acknowledges support from NSF MRSEC DMR05-20020. JFD acknowledges support from the NSF under DMR08-4376. Special thanks go to Prof. Jay Kikkawa and his students for experimental assistance.

**FIGURE CAPTIONS**

**Fig. 1** Film structure. X-ray diffraction from $Fe_{1-x}Co_xSi$ thin films grown via laser ablation with 900 s deposition time with *x* values denoted in the figure. Inset: Lattice constant versus Co concentration for films (bullets) and bulk samples (filled squares). Solid lines through data are a linear fit confirming Vegard's law.

**Fig. 2** (a) TEM image of FeSi thin film (15 minutes deposition time). (b) AFM image of $Fe_{0.85}Co_{0.15}Si$ thin film (10 minutes deposition time). The image is $10 \times 10 \mu m^2$.

**Fig. 3** Magnetic susceptibility. Magnetic susceptibility, $\chi$, versus temperature, *T*, in a magnetic field of 50 Oe for both zero-field-cooled (ZFC) and field-cool (FC) conditions for 3 $Fe_{1-x}Co_xSi$ thin films having a 900 s deposition time and a millimeter sized single crystal[23] as identified in the figure. Inset: Curie temperature, $T_C$, determined from the maximum of $\chi(T)$ for thin films and polycrystalline pellets[2] as identified in the figure

**Fig. 4** Magnetization. (a) Magnetization, *M*, at a field, *H*, of 10 kOe versus temperature, *T*, for the same $Fe_{1-x}Co_xSi$ thin films as in Fig. 3 and polycrystalline pellets with $x = 0.2$ and $x = 0.3$ (taken from Ref. 2) as identified in the figure.(b) *M* versus *H* at 5 K for $Fe_{1-x}Co_xSi$ films (15 minutes deposition time) with x = 0, 0.1, 0.15, and 0.2 and our arc-melted polycrystalline pellet with x = 0.15 (taken from Ref. 2) as identified in the figure. Arrows indicate the sequence of magnetic fields.

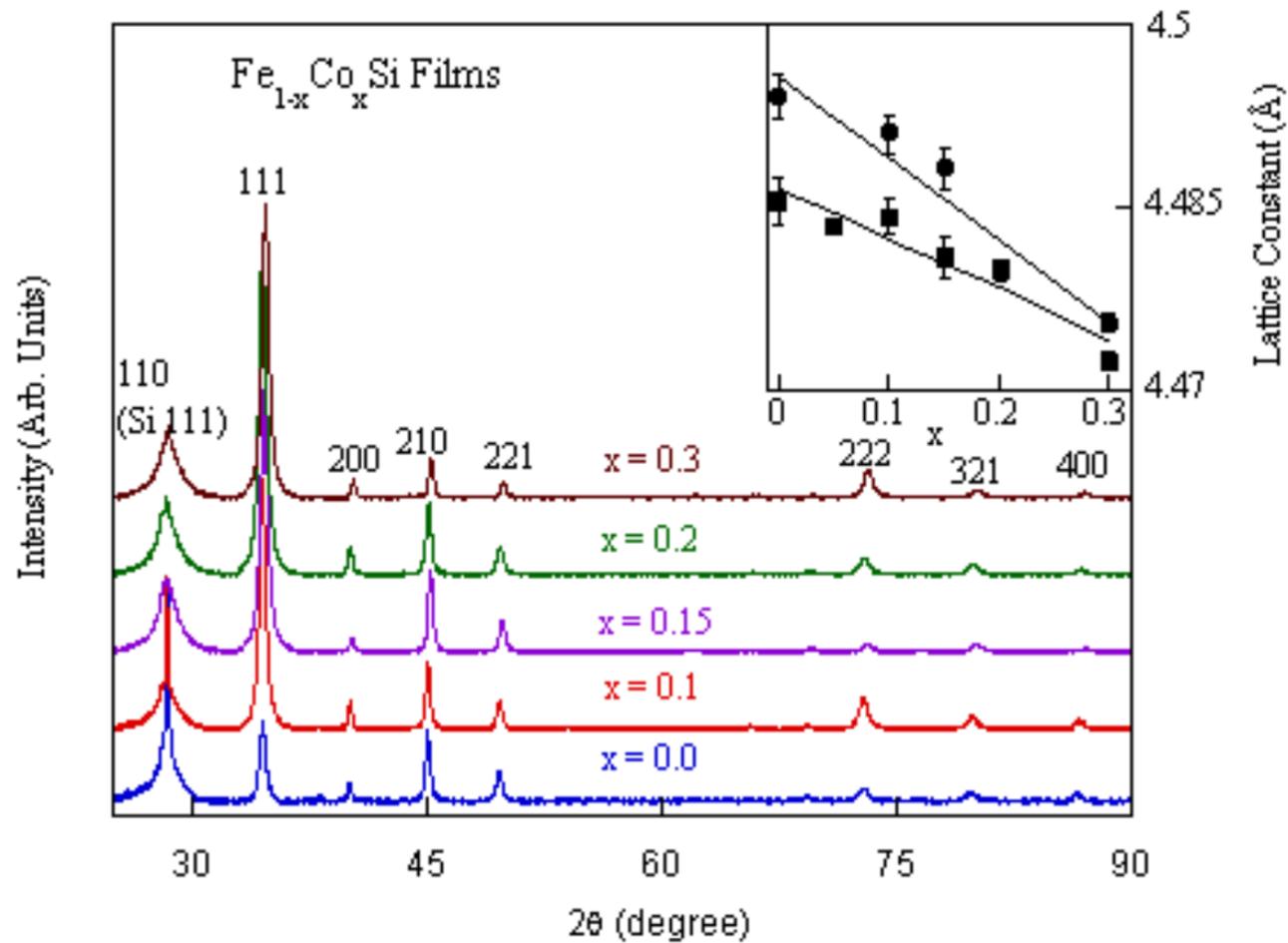

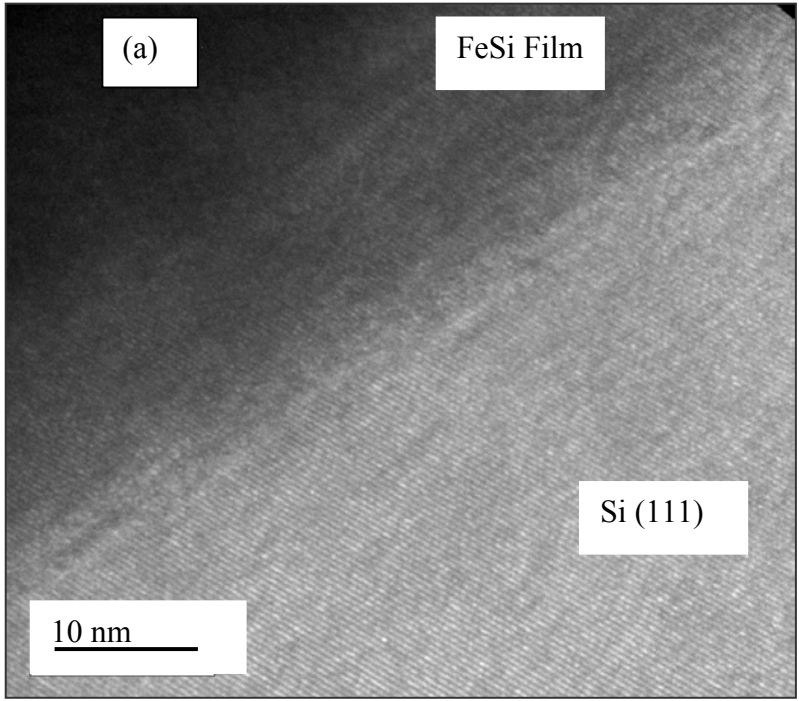

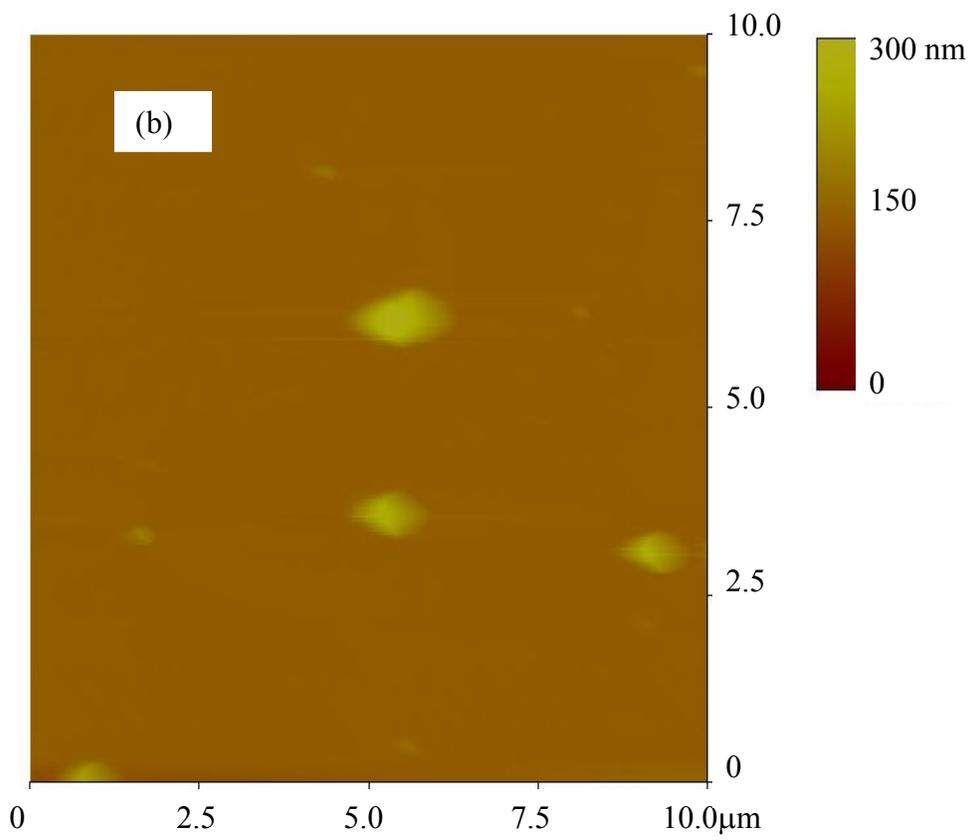

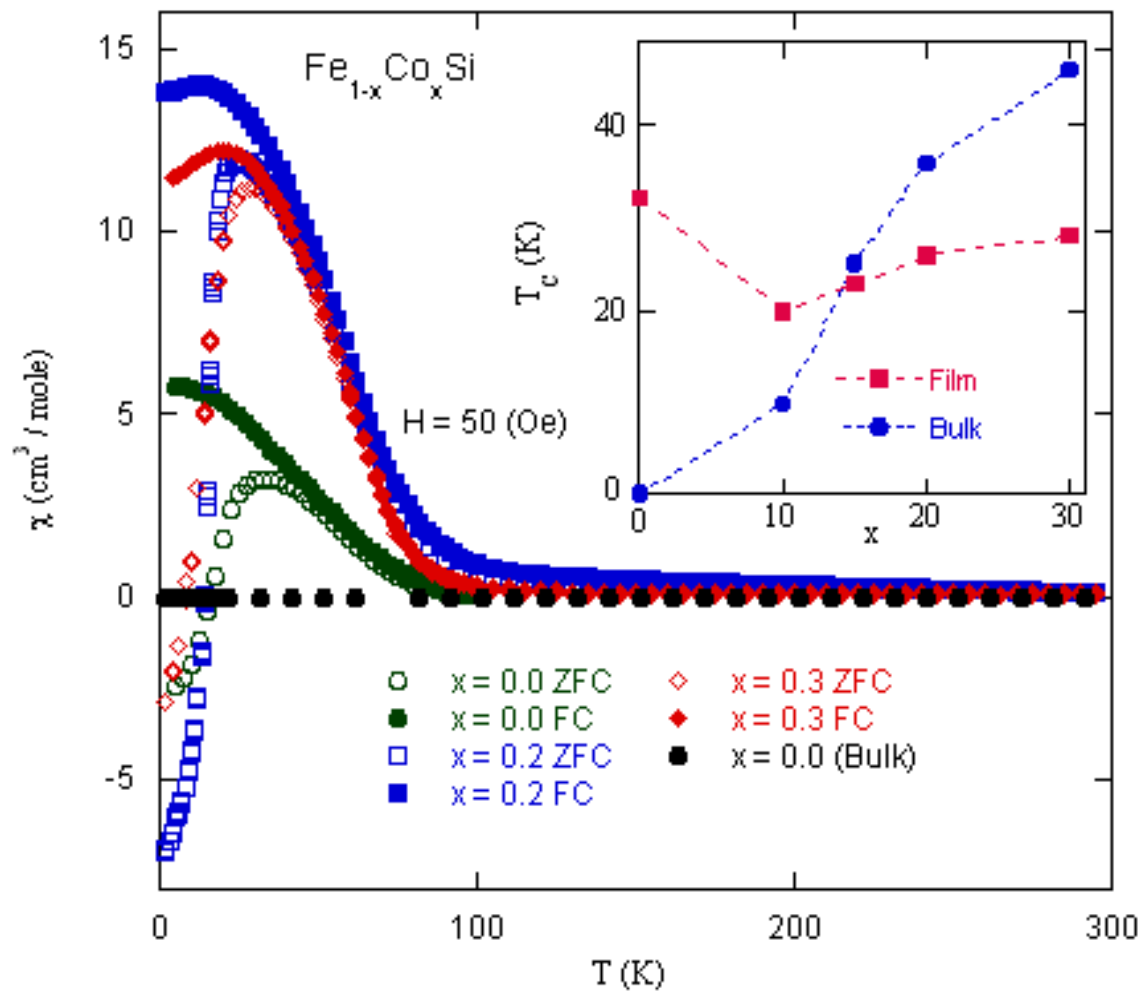

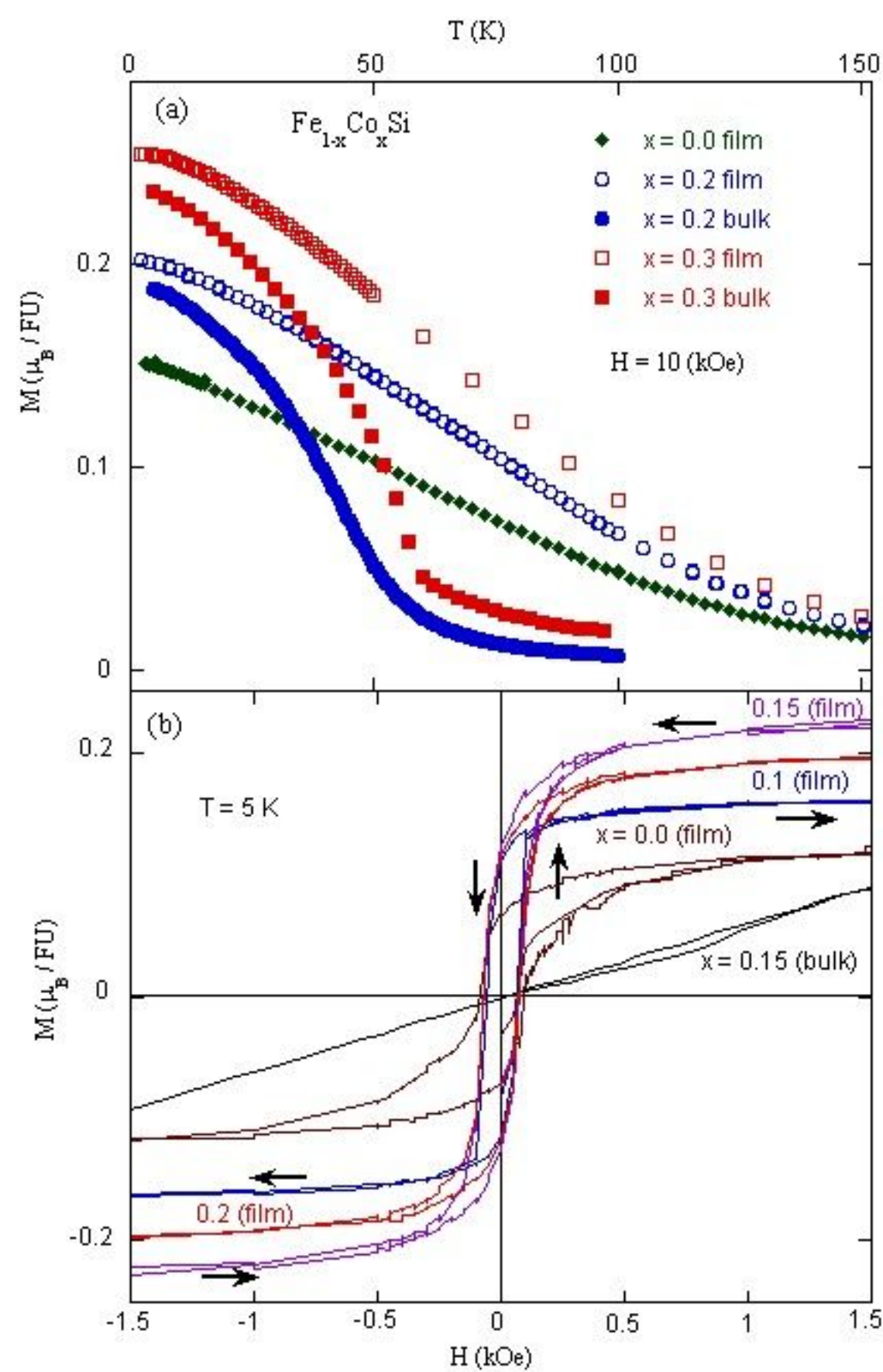